\documentclass[12pt]{elsart}
\usepackage{epsfig}

\def\hc#1{\leavevmode\hbox to\textwidth{\hss #1\hss}\leavevmode}

\date{3 June 1999}
\journal{Astroparticle Physics}

\begin{document}

\newif\ifForReferee \ForRefereefalse
\ifForReferee
\widowpenalty=9000
\clubpenalty=8000
\baselineskip=1.333\baselineskip
\makeatletter
\textfloatsep 12\p@ \@plus 4\p@ \@minus 2\p@ 
\def\small{\@setfontsize\small\@xipt{16}
\abovedisplayskip 11\p@ \@plus3\p@ minus6\p@
\belowdisplayskip \abovedisplayskip
\abovedisplayshortskip  \z@ \@plus3\p@
\belowdisplayshortskip  6.5\p@ \@plus3.5\p@ minus3\p@
\def\@listi{\leftmargin\leftmargini
 \parsep 4.5\p@ \@plus2\p@ minus\p@ \itemsep \parsep
            \topsep 9\p@ \@plus3\p@ minus5\p@}}
\makeatother
\fi

\begin{frontmatter}
\title{Impact of atmospheric parameters on
   the atmospheric Cherenkov technique%
\thanksref{titlecom}
}
\thanks[titlecom]{Based, in part, on work at University of Hamburg
 and Forschungszentrum Karlsruhe.}
\author{Konrad Bernl\"ohr}
\address{   Max-Planck-Institut f\"ur Kernphysik \\
   Postfach 103980, 69029 Heidelberg, Germany }


\begin{abstract}

Atmospheric density profiles as well as several light absorption and
scattering processes depend on geographic position and are
generally time-variable. Their impact on the atmospheric
Cherenkov technique in general (imaging or non-imaging) is
investigated. Different density profiles lead to differences
in Cherenkov light density of up to 60\%. Seasonal variations
at mid-latitude sites are of the order of 15--20\%. 
The quest for improved energy calibration of Cherenkov
experiments also shows the need for improved transmission
calculations, taking all relevant processes into account
and using realistic profiles of absorbers. Simulations including
the scattering mechanisms also reveal the relevance of
Rayleigh and Mie scattering for atmospheric Cherenkov experiments.
Refraction and the differences between treating the atmosphere
in plane-parallel or spherical geometry are also investigated.

\end{abstract}


\end{frontmatter}


\section { Introduction }

The atmospheric Cherenkov technique for air-shower detection, 
in particular the {\em imaging\/}
atmospheric Cherenkov technique, has become an increasingly mature 
experimental method of very-high-energy (VHE) $\gamma$-ray astronomy 
\cite{Weekes-1988,Cawley-1995,Aharonian-1997,Ong-1998}
in recent years. 
Large effort has gone, for example, into the optimisation of
$\gamma$-hadron separation, the energy calibration, and 
the evaluation of spectra of $\gamma$-ray sources. Imaging and non-imaging
methods play an increasingly important part in measuring the spectrum
and composition of cosmic rays, from TeV energies well into the knee
region. Among the non-imaging techniques, (former) solar power plants 
with dedicated Cherenkov equipment \cite{CELESTE,STACEE} have begun to achieve
unprecedentedly low energy threshold for ground-based $\gamma$-ray detectors.
The technique of imaging Cherenkov telescopes is also evolving, with
several stereoscopic arrays of ten-meter class telescopes now under 
development \cite{VERITAS,HESS}, 
even larger single telescopes \cite{MAGIC}, and some hope for further
progress in photon detection techniques.

One very important common aspect in all the variants of
the atmospheric Cherenkov technique is the atmosphere itself,
as the target medium for the VHE cosmic particles, as the emitter of
Cherenkov photons, and as the transport medium for those photons.

The present paper tries to further our understanding of
several important atmospheric parameters, mainly by extensive
numerical simulations. Among the parameters investigated are
the vertical profile of the atmosphere, the transmission and scattering
of Cherenkov light, the importance of spherical versus plane-parallel
geometry in shower simulations (or its insignificance, depending
on zenith angle), and the refraction of Cherenkov light. 
Since available tools were not really adequate for most of the questions
involved, the Cherenkov part of the CORSIKA \cite{CORSIKA} air shower
simulation program has been substantially extended and a flexible
and very detailed simulation procedure for imaging Cherenkov telescopes
developed (although the later is of less relevance for the present paper).

The major goal of this study is to be of practical usefulness for the
experimentalist. 


\section { Atmospheric profiles }

For the detection of air showers by particle detectors,
a pressure correction is usually sufficient to account for
different atmospheric density profiles. For the atmospheric
Cherenkov technique the situation is more complex since the
shower is not only sampled at one altitude but light is collected
from all altitudes. In addition to different longitudinal shower
development for different atmospheric density profiles, the
atmospheric Cherenkov technique is also sensitive to the
index of refraction $n$.
Both the amount of Cherenkov light emitted
and its emission angle are affected by the index of refraction
at each altitude.

The number of Cherenkov photons emitted per unit path length 
(in the wavelength range $\lambda_1$ to $\lambda_2$) is described
by the well-known equation
\begin{equation}
\frac{dN}{dx} = 2\pi\alpha z^2 \int_{\lambda_1}^{\lambda_2}
  \biggl(1 - \frac{1}{(\beta n(\lambda))^2}\biggr) 
  \frac{1}{\lambda^2} d\lambda,
\end{equation}
with $\alpha$ being the fine structure constant ($\approx$1/137),
$z$ being the charge number, and $\beta=v/c$. 
Particles with $\beta<1/n(\lambda)$ cannot emit Cherenkov light
at wavelength $\lambda$. At visible wavelengths this results
in an energy threshold of more than 20~MeV (35~MeV) for electrons
or positrons and about 4.5~GeV (8~GeV) for muons at sea level
(at 10~km altitude), respectively. The amount of light emitted
by particles above threshold depends
on the index of refraction.  The opening angle $\theta_{\rm c}$ of the 
Cherenkov light cone above threshold also depends on $n$:
\begin{equation}
   \cos \theta_{\rm c} = {1} / {n\beta}
\end{equation}
which in the limit $\beta=1$ and for $(n-1)\ll1$ corresponds to
\begin{equation}
   \theta_{\rm c} \approx \sqrt{2(n-1)} \quad\textrm{radians}.
\end{equation}

Different atmospheric density profiles, generally, result in
different indices of refraction near shower maximum and, thus, in
different amounts of Cherenkov light emitted. Most of the
light arriving in the inner region of fairly flat light
density (of about 120~m radius at 2000~m altitude) is emitted
near and after the shower maximum and is particularly affected 
by the longitudinal shower development as compared to
the total amount of Cherenkov light.

Since $n(\lambda)-1$ changes by only 5\% over
the wavelength range 300--600~nm, the range typically covered
by photomultipliers, air-shower Cherenkov simulations are
usually simplified by assuming a wavelength-independent index
of refraction $n$, obtained at an effective wavelength.
Another frequent simplification is to assume that $n-1$ is
proportional to air density but, strictly, $n$ is a more complex
function of pressure, temperature, and water vapour content.
For the present work, the wavelength independence is also
assumed but the dependence on water vapour content etc.\ is 
taken into account.

For the purpose of this study the CORSIKA shower simulation
program \cite{CORSIKA} has been adapted to read tables of
atmospheric profiles, including density and index of refraction,
and suitably interpolate between tabulated values. For the
electromagnetic part of the shower development, based on EGS
\cite{EGS}, several layers of exponential density profile are used
in CORSIKA. Fitting of the corresponding parameters to tabulated vertical
profiles can now be done at program start-up. The Cherenkov part of CORSIKA,
originally based on work of M.~Rozanska, S.~Martinez, and F.~Arqueros,
has been rewritten to account for tabulated indices of refraction
and for atmospheric refraction and includes an interface to
arbitrary systems of telescopes or other Cherenkov light detectors.

Seven atmospheric tables have been used for this work. Six tables were
obtained from the MODTRAN \cite{MODTRAN} program for atmospheric
transmission and radiance calculations (tropical, mid-latitude summer
and winter, subarctic summer and winter, and
U.S.\ standard atmosphere 1976). An antarctic winter profile
was constructed from radio\-sonde measurements above the Amundsen-Scott
(south pole, 2800~m altitude) and Neumayer (latitude 70$^\circ$~S,
near sea level) stations.

\begin{figure}
\epsfig{file=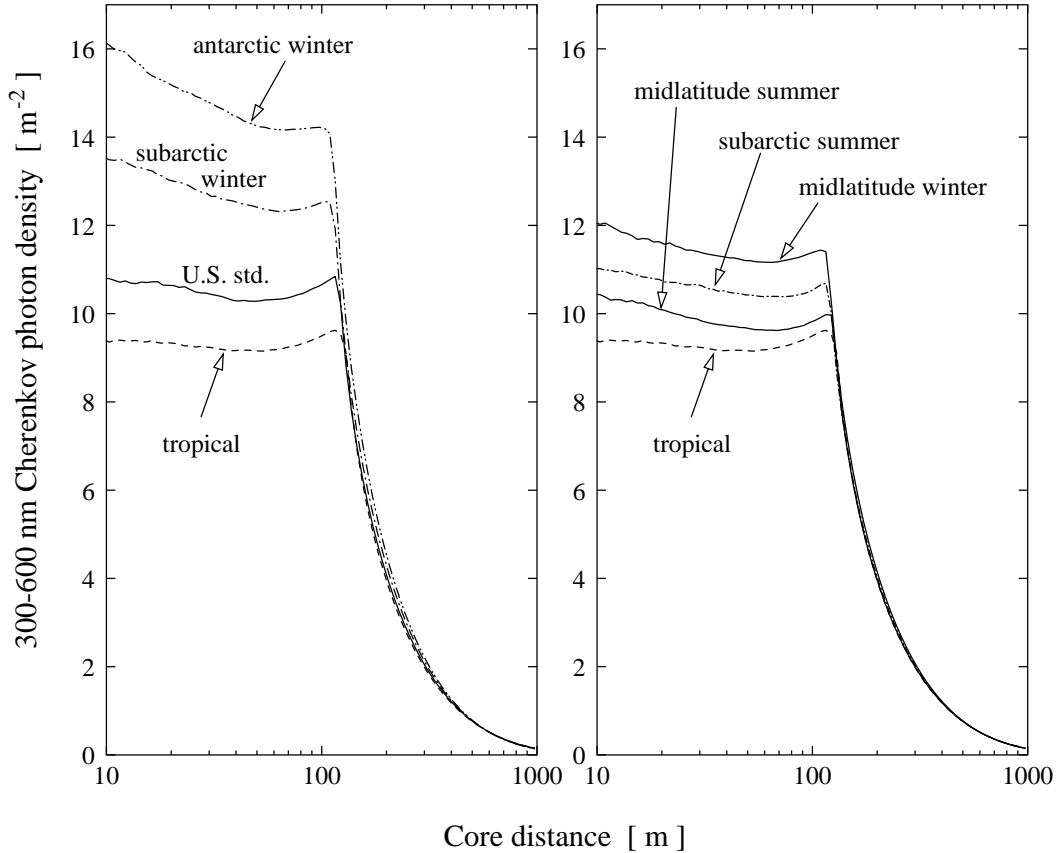,width=\hsize}
\caption[Lateral distributions of Cherenkov light]{Average lateral 
distributions of Cherenkov light photons in the
wavelength range 300--600~nm for vertical 100~GeV gamma-ray showers in
CORSIKA 5.71 simulations with different atmospheric profiles
(2000 showers simulated for each profile). 
Absorption of Cherenkov
light is taken into account (see section~\ref{sec:transmission}).
Observation altitude is 2200~m above sea level.}
\label{fig:lat-prof-1}
\end{figure}

Figure~\ref{fig:lat-prof-1} shows the quite significant impact
of the various atmospheric profiles on the lateral density of
Cherenkov light in 100~GeV gamma-ray showers observed
at an altitude of 2200~m (the impact being similar for any altitude far 
beyond shower maximum). The same atmospheric light transmission model is
used in all cases (see section~\ref{sec:transmission}).
The energy of 100~GeV has been chosen here because it will be a rather
typical energy for the next generation of atmospheric Cherenkov
experiments and because enough showers can easily be simulated
such that shower fluctuations cancel out to a negligible level.

A 60\% higher light density near the shower axis is obtained for the
antarctic winter as compared to the tropical profile. At moderate
latitudes a seasonal effect of 15--20\% is apparent and should
be included in energy calibrations of IACT installations.
Air-shower Cherenkov simulations used for the energy calibration
of various experiments have to date mainly used the U.S. Standard
Atmosphere 1976 \cite{US-Std-1976} profile. 
Inappropriate atmospheric models could lead to systematic errors in 
absolute flux calibrations of the Crab nebula -- 
the de-facto {\em standard\/} VHE $\gamma$-ray source. 

Flux calibrations relative to the
(not very accurately known) flux of cosmic rays would be less
subject to assumed atmospheric profiles. 
These relative flux calibrations are most useful for comparison between
different experiments but require that the same cosmic-ray flux and
composition are assumed.
The relative method also depends on applied hadronic interaction models 
which are still less accurate than electromagnetic shower codes even
at energies where accelerator data are available. 
Note also that, among absolute calibration methods for Cherenkov telescopes, 
both calibration with a reference light source and with muon rings
require detailed knowledge of the spectral response curve since neither
the light source nor the muon rings have the same spectrum as the 
Cherenkov light from near the shower maximum. To a lesser extent this
is also true for the relative method because hadron showers with
a deeper shower maximum and some light from penetrating muons have,
on average, less short-wavelength extinction than gamma showers.

The atmospheric profile is not only important for
the average light density at small core distances but also for
the radial fall-off.  At multi-TeV energies, this radial fall-off 
is useful as a means to discriminate between
hadron and gamma-ray initiated showers and 
to estimate the cosmic-ray mass composition.
Simulations with inappropriate atmospheric profiles could lead 
to systematics in both cases.

\begin{figure}
\epsfig{file=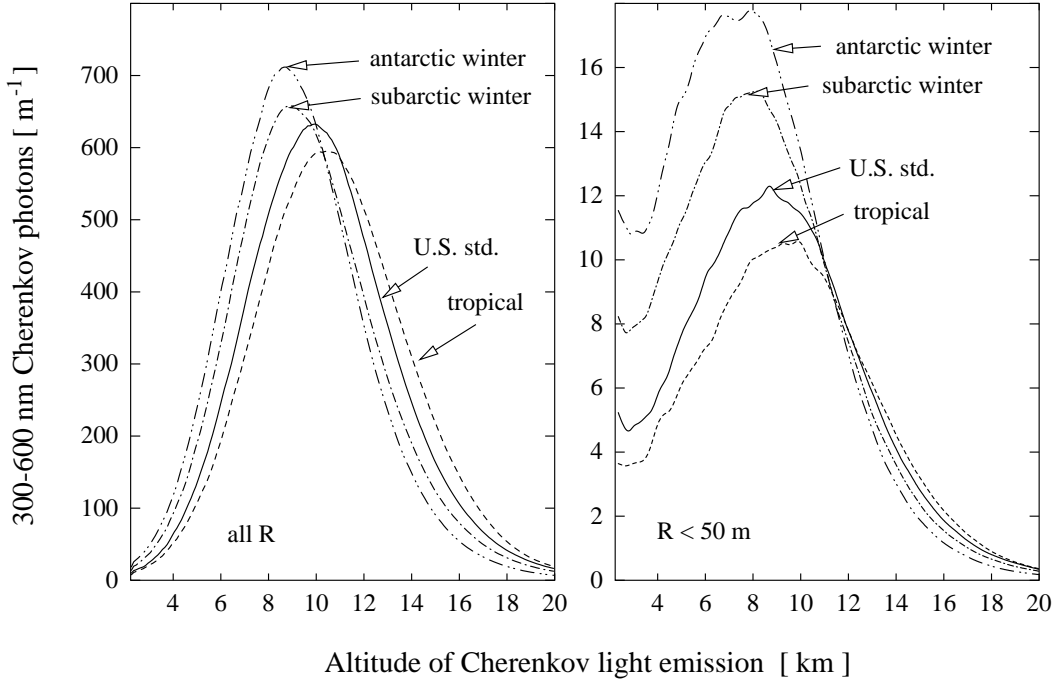,width=\hsize}
\caption[]{Average Cherenkov light emission along the
 shower axis for vertical 100~GeV gamma-rays with different
 atmospheric profiles. Left: All emitted photons. Right:
 Photons which would arrive within 50~m from the core at the
 observation level of 2200~m. No absorption is applied here.}
\label{fig:long-prof-1}
\end{figure}

The reasons for the different light profiles are illustrated to some 
extent by Figure~\ref{fig:long-prof-1}, showing the average longitudinal 
development of showers for four profiles. For profiles with
lower temperatures in the lower stratosphere and troposphere
the maximum of Cherenkov emission is shifted downwards --
to regions of higher density, i.e.\ higher index of refraction
and thus higher Cherenkov efficiency -- with respect to
profiles with higher temperatures. 
It should be noted that the atmospheric thickness corresponding
to the height of maximum of all Cherenkov emission remains largely unaffected
(not more than 5~g/cm$^2$), but the thickness of the maximum of
emission into the inner 50~m is increasing substantially from
the tropical to the antarctic winter profile (by about 30~g/cm$^2$).

The amount of Cherenkov
light within 500~m from the core is roughly proportional to
$(n-1)$ at the shower maximum, with about 15\% difference between
tropical and antarctic winter. If light arriving very far from
the shower core is included, the differences are even smaller.
Near the core, however, differences are large 
(see Figure~\ref{fig:lat-prof-1}) which is due to several
effects:
\begin{itemize}
\item The amount of Cherenkov emission is roughly proportional
   to $(n-1)$ at median altitude $h_{\rm med}$ of Cherenkov
   emission (or at maximum, as before).
\item With increasing $(n_{\rm med}-1)$ at $h_{\rm med}$ the Cherenkov 
   cone opening angle is increased and the light is spread
   over a larger area -- decreasing the central light density.
\item With decreasing $h_{\rm med}$ the distance between
   emission maximum and observer is decreased -- increasing 
   the central light density.
\item For Cherenkov light near the core, the median height of emission
   $h_{\rm{med}}^\star$
   is typically 1000--1500~m below that of all Cherenkov light 
   ($h_{\rm{med}}$), which
   emphasises the geometrical factor even more.
\end{itemize}
Qualitatively, the central light density $\rho_{\rm c}$  for
vertical showers follows
\begin{equation}
   \rho_{\rm c} \propto \frac{n_{\rm{med}}-1} 
      {(n_{\rm{med}}-1) \,(h_{\rm{med}}^\star-h_{\rm{obs}})^2}
      = (h_{\rm{med}}^\star-h_{\rm{obs}})^{-2}
\end{equation}
where $h_{\rm{obs}}$ is the observation level altitude. The numerator
accounts for the Cherenkov efficiency, the denominator for the
area of the light pool. The index of refraction cancels
out in this approximation, 
leaving the distance between $h_{\rm{med}}^\star$ and $h_{\rm{obs}}$
as the dominating factor.
Since for increasing primary energy $h_{\rm{med}}$ approaches 
the observation altitude, the geometrical
factor will be increasingly important.

\begin{figure}
\epsfig{file=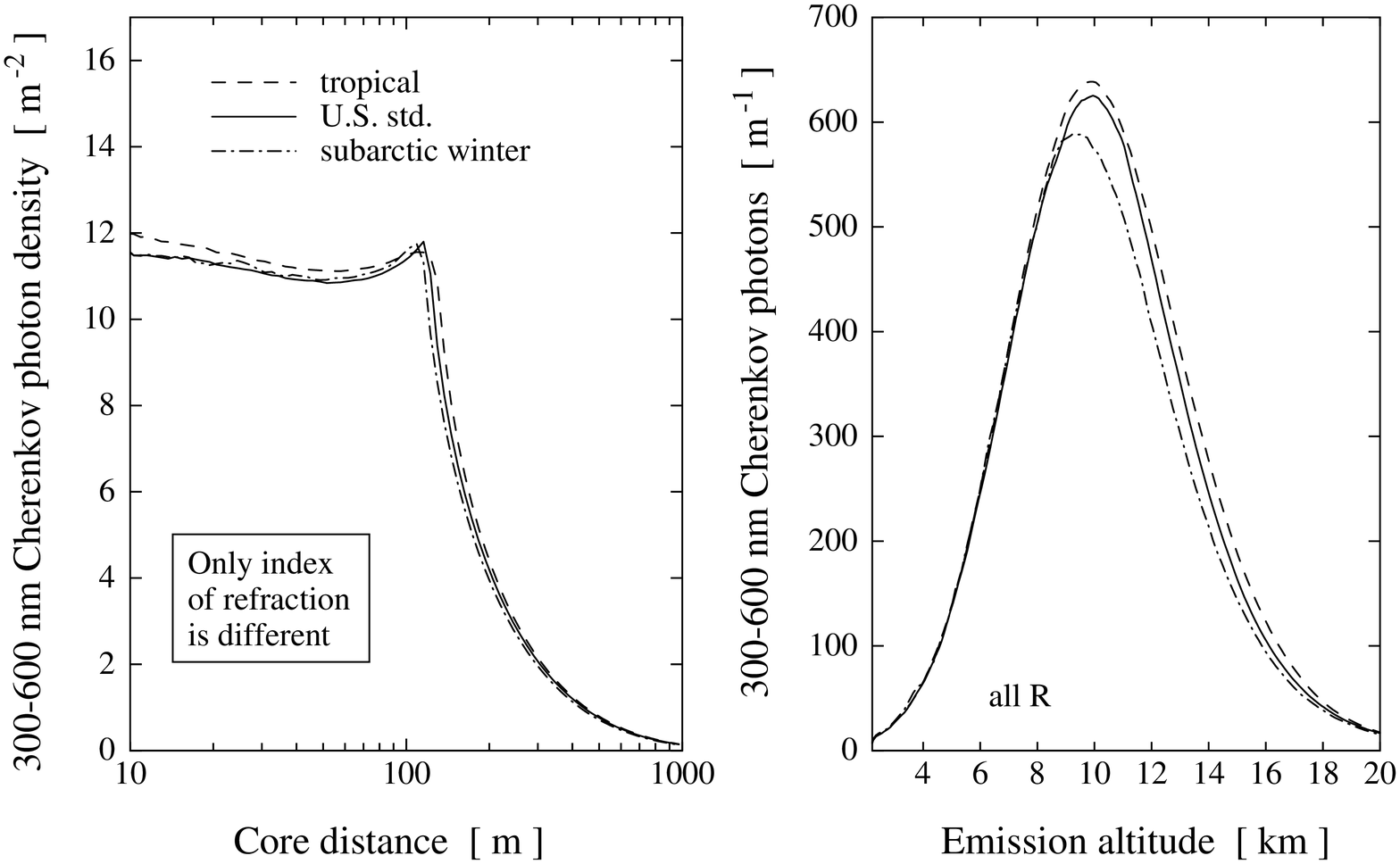,width=\hsize}
\caption[Cherenkov light with different index of refraction profiles]{Cherenkov
light profiles when only the index of refraction is taken from different
atmospheric profiles but shower development is in all cases simulated with
U.S. standard atmosphere.
Left: Lateral density of Cherenkov photons (as in Figure~\ref{fig:lat-prof-1}).
Right: Longitudinal profile of Cherenkov emission 
(as in the left panel of Figure~\ref{fig:long-prof-1}).}
\label{fig:lat-prof-2}
\end{figure}

In simulations, it is possible to separate the effects of the
atmospheric profiles on shower development and on Cherenkov emission.
In Figure~\ref{fig:lat-prof-2} the shower development is treated 
with the CORSIKA built-in U.S. standard atmosphere approximation
but the index of refraction is taken from different atmospheric profiles.
Since the impact of the different distances between observation level
and median emission altitude is not present in this case, any
differences in lateral light density are much smaller.
The position and shape of the
rim of the `light pool' 100-120~m from the core are the most
obvious differences remaining, which are due to
different Cherenkov cone opening angles in the lower stratosphere.


\section { Transmission of Cherenkov light }
\label{sec:transmission}

\begin{figure}
\hc{\epsfig{file=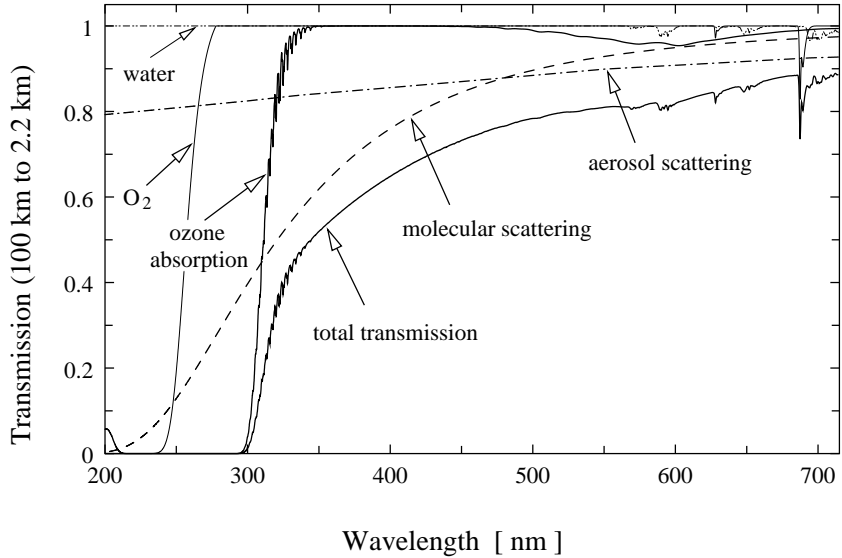,width=0.8\hsize}}
\caption[Transmission by components]{Direct transmission of light
from space (here 100~km altitude) along a vertical path to an
altitude of 2.2~km, as calculated with MODTRAN. 
The impact of the most important absorbers
and scatterers is shown.}
\label{fig:trans1}
\end{figure}

The atmospheric extinction of light is another source of concern for the
energy calibration of atmospheric Cherenkov
experiments and to some extent also for the image parameters of
telescopes.
There are several sources of extinction: absorption bands of several
molecules, molecular (Rayleigh) scattering as well as aerosol 
(Mie) scattering and absorption. 
For a detailed introduction see for example
\cite{Kyle-1991}. The relevance of the various absorbers or scatterers
at different wavelengths is illustrated in Figure~\ref{fig:trans1}.

At wavelengths below 340~nm ozone (O$_3$) is a very important absorber --
not only in the ozone layer but even near ground. Relevant absorption
bands are the Hartley bands in the 200-300~nm range and the Huggins bands
extending to 340~nm. Near 600~nm there are the weak Chappuis bands.
Normal oxygen (O$_2$) can be disassociated by light below
242~nm leading to the Hertzberg continuum. In addition, there is the
Hertzberg band at 260~nm. The O$_2$ absorption is of no concern to
most Cherenkov experiments -- typically using photomultipliers (PMs) with
borosilicate glass windows which are insensitive below 290~nm -- 
and is in fact frequently neglected.
However, O$_2$ absorption is a limiting factor for UV
observations. Other molecules are of little relevance in the 
near-UV and visible range.

Most Cherenkov light in the PM sensitivity range is actually lost by
molecular scattering. Although some of the light may also be scattered
into the viewing angle, such scattered light
is generally not important and scattering can be considered like
an absorption process. The same argument applies to aerosols where
both scattering and absorption play a role. The relevance of scattered
light is discussed in Section~\ref{sec:scattering}.

While molecular scattering and O$_2$ absorption are easily predictable
and almost constant at any site, both aerosols and ozone are site-dependent 
and variable. Aerosols are mainly limited to the {\em boundary layer} of
typically 1--2.5~km thickness above the surrounding terrain where the
diurnal variation and the dependence on ground material and wind speed
is largest. In the boundary layer, the heating of the ground by solar
radiation leads to turbulence and rapid vertical exchange of air
and dust. 
Not just near ground but even in the stratosphere the aerosols play a role --
including meteoric and volcanic dust. Ozone also shows diurnal and
seasonal variations. 

The total extinction of star light is easily measured
(and a routine procedure at optical observatories), by fitting
the function
\begin{equation}
\ln I(\lambda) = \ln I_0(\lambda) \, - \tau_1(\lambda) \sec z
\end{equation}
to several observations of a reference star (here in
the plane-parallel atmosphere approximation).
In this equation $I$ is the measured intensity, $I_0$ the true intensity,
$\tau_1$ the optical depth per unit airmass and $\sec z$ the
secant of the zenith angle. For the procedure one or several
sources are measured at widely different zenith angles, allowing
to fit $I_0$ and $\tau_1$.
This procedure, however, cannot disentangle the vertical structure
of absorbers. Different assumptions on this structure easily
lead to differences of 5--10\% in the amount of Cherenkov light,
even at mountain altitude. At sea-level, even differences of up to
30\% between different calculations can be traced back to
different assumptions on the extinction. 

\begin{figure}
\hc{\epsfig{file=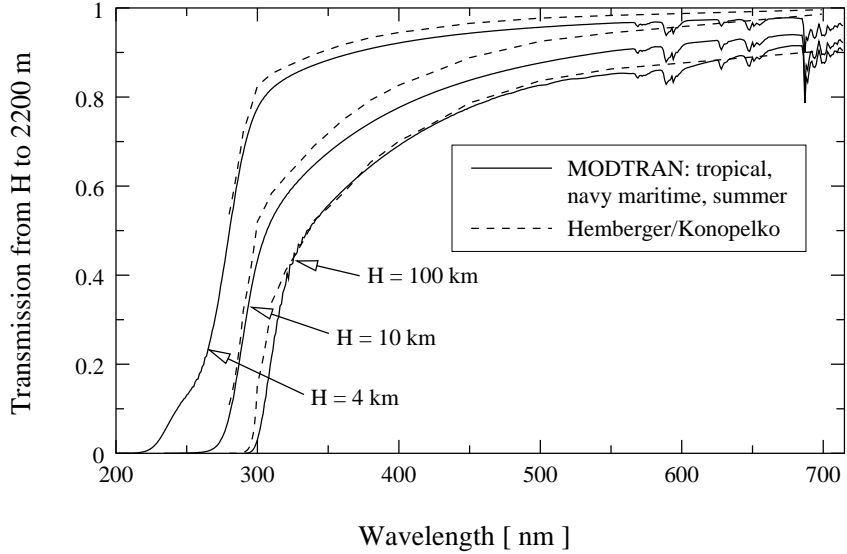,width=0.8\hsize}}
\caption[Transmission with two different aerosol models]{Comparison
of atmospheric transmission as calculated with MODTRAN (using the
tropical profile and navy maritime summer haze model) and in a
transmission model with aerosol absorption proportional to
atmospheric density \cite{Konopelko-xxxx,Hemberger-priv-1997}.
Note that although both transmission models have almost the
same transmission for stellar light, the transmission from typical
Cherenkov emission altitudes differs significantly.}
\label{fig:trans2}
\end{figure}

One example of a bad
assumption is to take the density of aerosols as proportional to
air density. One such example is illustrated in
Figure~\ref{fig:trans2}. The aerosol-air proportionality 
assumption leads to an over-estimate (by 4--8\%) of Cherenkov light
even if the measured star-light extinction at the actual (mountain) altitude
is taken into account. The reason for that is that the
Cherenkov light is produced, say, halfway down in the atmosphere,
implying 50\% of the star-light extinction under the assumption,
but actually some 80--90\% of the aerosol extinction happens below the
average Cherenkov production altitude.
The aerosol-density proportionality assumption together with the 
extrapolation of mountain-altitude extinction measurements down to sea level,
for example, leads to a severe over-estimate of Cherenkov light
intensity at sea level. 

\begin{figure}
\hc{\epsfig{file=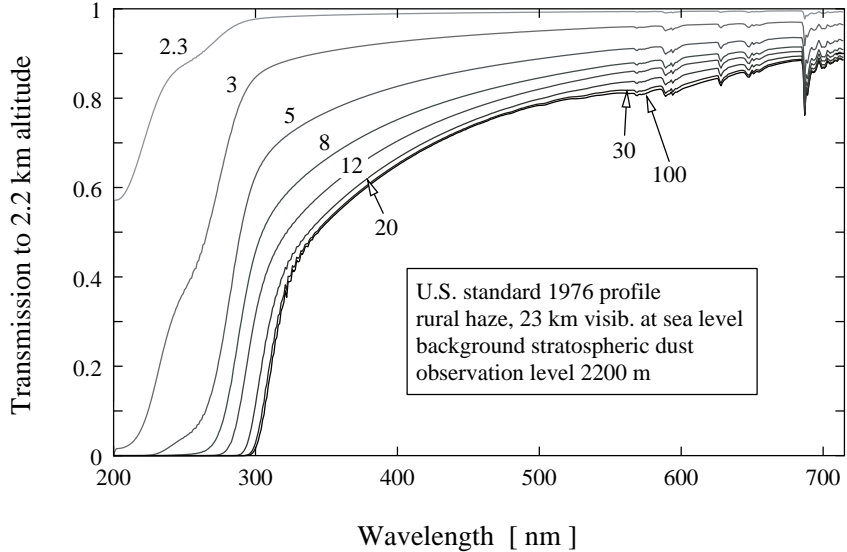,width=0.8\hsize}}
\caption[Transmission curves from different altitudes]{Direct transmission
for light along vertical paths from different altitudes (in km),
as noted in the figure,
to an observation level of 2200~m. Transmission was calculated with
MODTRAN for U.S. standard atmospheric profile, rural haze with
23~km sea level horizontal visibility, and background stratospheric
dust.}
\label{fig:trans3}
\end{figure}

A much more realistic model of aerosol vertical structure,
aerosol properties plus all the relevant molecular absorption
and scattering is included in the MODTRAN \cite{MODTRAN} program.
MODTRAN has been used for the extinction models used in this paper.
Unless otherwise noted, a U.S. standard profile with rural haze model 
of 23~km sea-level horizontal visibility has been used. Transmission
curves obtained with this model are shown in Figure~\ref{fig:trans3}.

\begin{figure}
\hc{\epsfig{file=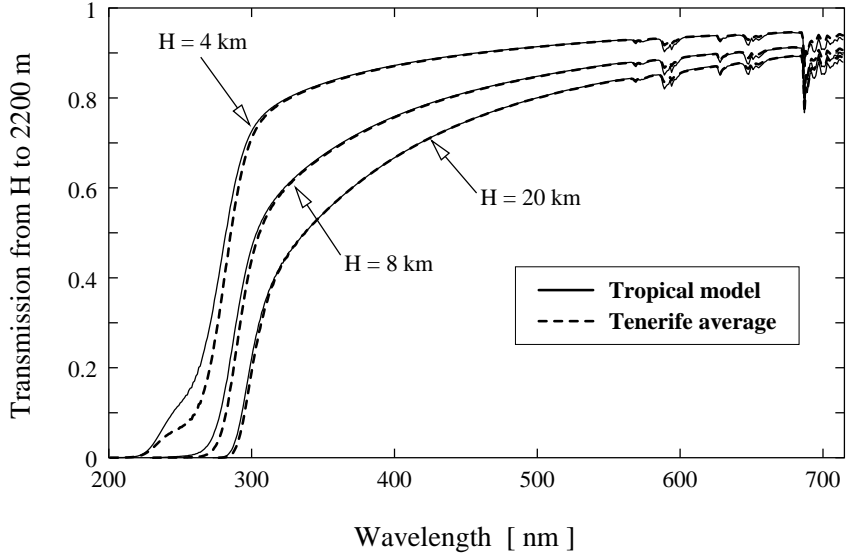,width=0.8\hsize}}
\caption[Impact of tropospheric ozone variation]{Impact of the
1.5--2 fold higher ozone content near ground measured 
\cite{Tenerife-ozone} with
radiosondes from Tenerife, Canary Islands, as compared to the
tropical profile. Transmissions calculated with MODTRAN.
For the Tenerife model the all-year average data was used. Above the
altitude reached by radiosondes (about 32~km), the tropical profile
is used. The same maritime haze model is used in both cases.}
\label{fig:trans4}
\end{figure}

In the case of the observatory on La Palma (operated
by the IAC), the tropical profile and navy maritime haze model
 -- with quite little aerosol extinction -- 
is in excellent agreement with measured extinction curves
as well as with the long-term average V band extinction, as
provided by the Isacc Newton group on La Palma and the
Royal Greenwich Observatory, Cambridge.
Temperature and pressure profiles of the MODTRAN tropical profile
are also in quite good agreement with radiosonde measurements from
the nearby Tenerife island~\cite{Tenerife-ozone}. 
Tropospheric ozone measurements (with the
same radiosondes), however, exceed the MODTRAN model by a
factor of 1.5--2. In Figure~\ref{fig:trans4} the transmission curves 
obtained with the built-in tropical profile and with the profile
taken from the radiosonde measurements are compared. For Cherenkov
measurements with borosilicate window PMs the differences are
insignificant but for UV observations they are important.

\begin{figure}
\hc{\epsfig{file=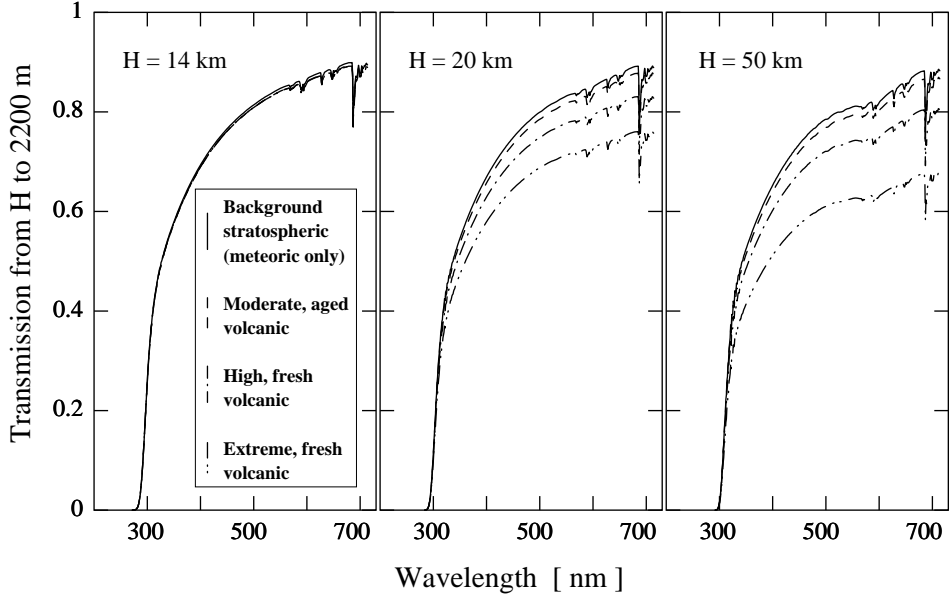,width=0.9\hsize}}
\caption[Transmission due to volcanic dust]{Atmospheric transmission for
different levels of volcanic dust extinction in the stratosphere.
The U.S. standard atmospheric profile is used with the rural haze model for 
tropospheric aerosols and meteoric plus volcanic dust in the stratosphere.
The Pinatubo eruption corresponds to activity slightly
exceeding the {\em high\/} scenario.}
\label{fig:trans5}
\end{figure}

As a further atmospheric variable the impact of volcanic dust was
studied. The 1991 Pinatubo eruption, for example, led to
30 million tons of stratospheric dust -- compared to 1 million tons
before the eruption. It is visible in
La Palma extinction measurements (obtained with the Carlsberg
Meridian Circle and made available by the Royal Greenwich Observatory, 
Cambridge) for a period of two years. This eruption had a high
(5--10\%) impact on extinction of stellar light for one and a half
years. MODTRAN provides a set of options to enhance the amount of
volcanic dust in the calculations. Results for different amounts of dust
and different Cherenkov emission altitudes are shown in
Figure~\ref{fig:trans5}. The volcanic dust extinction is insignificant
for altitudes below about 14~km and as such has little impact on
Cherenkov measurements. A calibration of
Cherenkov telescopes with stellar light under high volcanic dust conditions
could lead to an over-estimate of shower energies and, thus, fluxes.

\begin{figure}
\hc{\epsfig{file=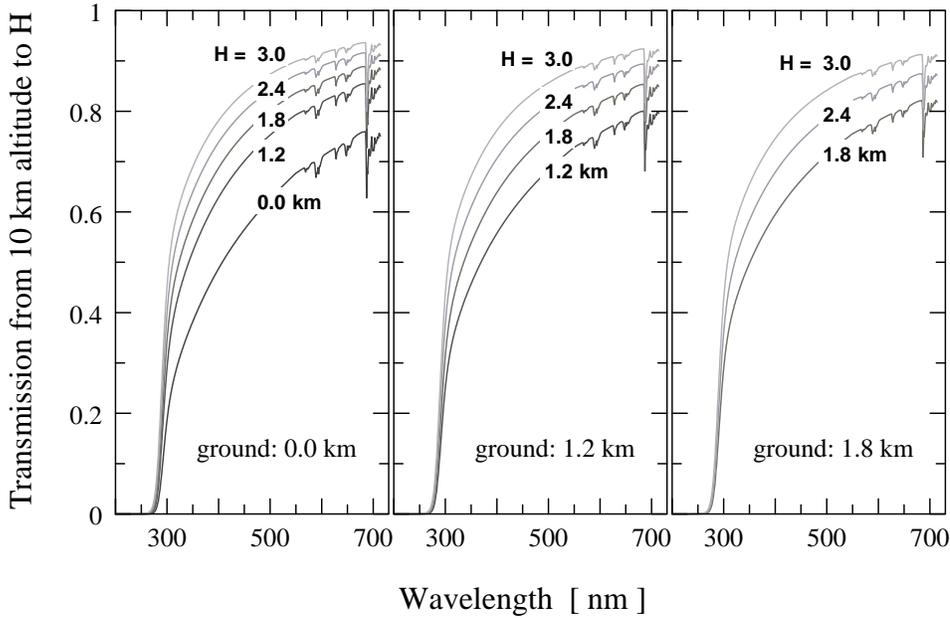,width=0.9\textwidth}}
\caption[Atmospheric transmission at different altitudes]{Atmospheric
transmission from 10~km to different observation levels (3.0, 2.4, 1.8, 1.2,
and 0.0 km a.s.l.), for three different `ground' altitudes (left: sea level,
middle 1.2~km, right: 1.8~km) as the base of the boundary layer.
MODTRAN calculations with U.S. standard atmosphere and rural haze model.}
\label{fig:trans-altitude}
\end{figure}

Some of the forthcoming Cherenkov installations \cite{VERITAS,HESS}
will likely be installed at the base of a mountain instead of at the top --
due to environmental or infrastructure reasons. It
seems appropriate to compare the expected atmospheric transmission
for sites at different altitudes. Since the aerosol absorption is
strongest in the boundary layer, the altitude of the surroundings
(on a scale of the order of hundred kilometers and more) is also relevant.
If the `base' of the mountain is still above the boundary layer,
the reduced altitude should not affect the transmission very much.
If the base is already at the bottom of the boundary layer, even
a move from 2.4~km to 1.8~km altitude results in 10\% less Cherenkov
light (Figure~\ref{fig:trans-altitude}, as deduced with the 
MODTRAN rural haze model). These calculations still
assume clear nights while in practice the base of the mountain
may be more frequently under a cloud layer or affected by ground fog
than the top. This, of course, can only be resolved by a long-term
site comparison.


\section { Scattering of light }
\label{sec:scattering}

In the preceding discussion, all molecular and aerosol scattering of
Cherenkov light is treated as an absorption process. This assumption
was apparently used in any atmospheric Cherenkov simulations so far.
However, estimates of the impact of scattered Cherenkov light were
taken into account for the fluorescence technique \cite{FlysEye}.
In this section quantitative results of full Cherenkov simulations with
scattered light are presented.\footnote{Since fluorescence light is not 
available with CORSIKA yet, these simulations could not be applied to the
fluorescence technique at this stage.}

When considering scattered light one has to take the relevant
integration time into account. Hardly any scattered light will 
arrive within or even before the Cherenkov light shower front but
most scattered light arrives with quite significant delay due to its detour.
For short integration times, small-angle scattering is responsible
for most of the scattered light. 

Rayleigh scattering (of
unpolarised light) is described by the simple normalised 
{\em phase function} of scattering angle $\gamma$
\begin{equation}
P_{\rm R}(\gamma)=\frac{3}{16\pi}\frac{2}{2+\delta} \,
 \Bigl((1+\delta)+(1-\delta)\cos^2\gamma \Bigr)
\end{equation}
with $\delta$ being the depolarisation factor due to anisotropic
molecules ($\delta\approx0.029$). For aerosol scattering -- in principle
described by Mie scattering theory -- the situation is much more complicated
and depends on size distribution, composition, and shapes of aerosol
particles. In all practical cases, aerosol scattering is quite
asymmetric with a forward peak. Due to its forward peak, aerosol
scattering generally dominates over molecular scattering in
Cherenkov light measurements with short integration times.

\begin{figure}
\hc{\psfig{file=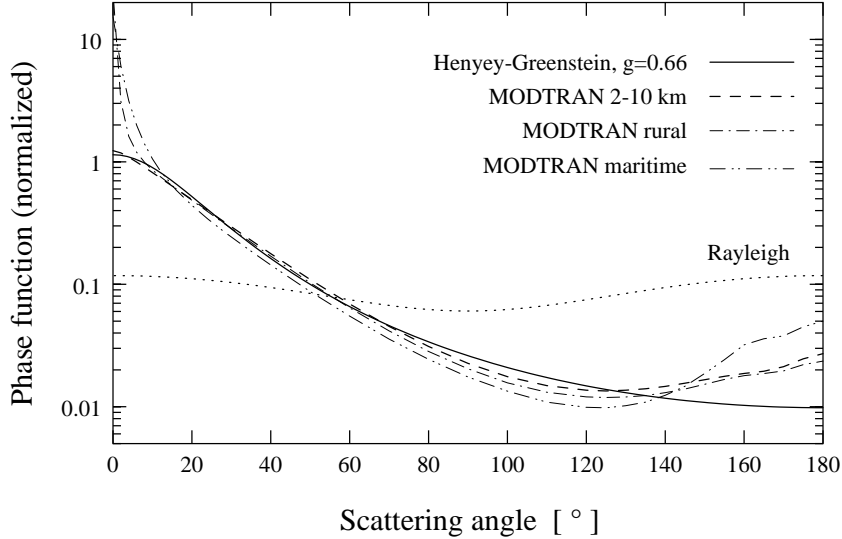,width=0.8\textwidth}}
\caption[Phase functions]{Phase functions, i.e.\ relative intensity
of scattered light per unit solid angle, for Rayleigh scattering (dotted),
the MODTRAN \cite{MODTRAN} 2--10~km tropospheric aerosol model (dashed) and the
Henyey-Greenstein function of $g=0.66$ (solid line). Also shown:
MODTRAN phase functions for rural and maritime 0--2~km boundary layer 
(thin dot-dashed lines).}
\label{fig:phase-functions}
\end{figure}

Although Cherenkov light of air showers is partially polarised, the
polarisation is ignored in the following because it is only relevant
for large-angle scattering. It should also be noted that the amount
of aerosol scattering (and to some extent also its phase function)
can be highly variable -- a fact that is very important for the
air shower fluorescence technique where the contamination of the weak
fluorescence light by scattered (in addition to direct) Cherenkov light 
has to be (and usually is) taken into account.
In the following, an average amount and phase function
for aerosol scattering is assumed which should be more or less typical
for a good astronomical site situated well above the boundary layer
in which turbulent mixing due to the diurnal temperature cycle is relevant.

Aerosol scattering and absorption coefficients have been calculated with
the MODTRAN \cite{MODTRAN} program. The phase function can be approximated
by a Henyey-Greenstein phase function with asymmetry parameter $g$:
\begin{equation}
P_{\rm HG}(\gamma) = \frac{1}{4\pi}\frac{(1-g^2)}{(1-2g\cos\gamma+g^2)^{3/2}}.
\end{equation}
The tropospheric aerosol phase function in MODTRAN has an 
asymmetry $g=\langle \cos\gamma P_{\rm HG}(\gamma)\rangle\approx0.7$
and  is in the angle range $0^\circ<\gamma<140^\circ$ well represented
by a Henyey-Greenstein phase function of $g\approx0.66$ 
(see Figure~\ref{fig:phase-functions}). 
In the following, $g=0.7$ is used. For the shower simulations with
a modified CORSIKA 5.70 program an 
U.S.\ standard atmospheric profile was used unless otherwise noted. 
Atmospheric transmission
coefficients (including absorption and scattering on aerosols)
were used as calculated with the MODTRAN rural haze model.
The scattering algorithm used with CORSIKA includes multiple scatterings
although these turned out to be insignificant.
An observation level at an altitude of 2200~m is assumed.
Since the relevance of scattered light is wavelength dependent,
usual observation conditions are simulated by applying
the quantum efficiency curve of a photomultiplier (PM) with 
borosilicate glass window and bi-alkali photocathode and 
the reflectivity of an aluminised mirror.

\begin{figure}
\hc{\psfig{file=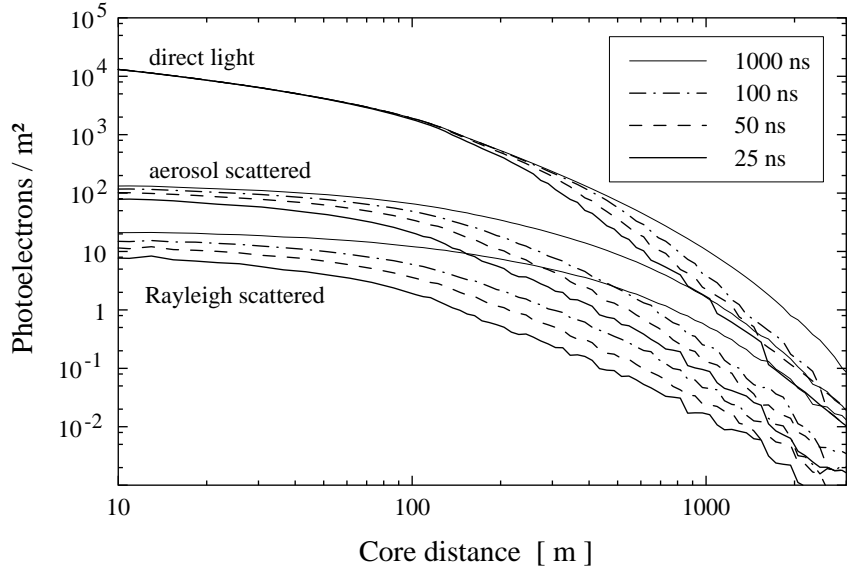,width=0.8\textwidth}}
\caption[Scattered light in 100 TeV proton showers]{Lateral density of
direct, aerosol scattered, and Rayleigh scattered light in 
vertical proton showers of 100~TeV primary energy (average of
10 showers) for different integration times. 
Note that integration
intervals at different core distances are shifted approximately such as 
to maximise the integrated average signal at each distance interval. 
Mirror reflectivity and PM quantum efficiency curve are applied for
the conversion from photons to photo-electrons.
A tropical atmospheric profile was used in this case.}
\label{fig:proton-scatter}
\end{figure}

The relevance of scattered light integrated over the whole sky is shown
in Figure~\ref{fig:proton-scatter} for 100~TeV proton showers. Note that
within the central 200~m, 1--3\% of the total Cherenkov light is scattered
light (for integration times below 100~ns). 
This fraction is increasing with distance since the lateral
distribution of scattered light is flatter than that of the direct light.
Within the central kilometer, aerosol scattered light dominates
over Rayleigh scattered light.
Beyond a few kilometers from the core and for integration times of more than
one microsecond, scattered light eventually exceeds
the direct light. Note that, for the short integration times, 
the smaller field of view of 
non-imaging Cherenkov counters like AIROBICC \cite{AIROBICC} or 
even BLANCA \cite{BLANCA} will not
much reduce the amount of scattered light. Most of the
scattered light arriving within a few 10~ns of the direct light
is scattered by no more than ten degrees.

In the imaging atmospheric Cherenkov technique the Cherenkov light
is integrated over a much smaller field of view and, generally, over
a very short time interval (20~ns or even less). In this case the
impact of scattered light is even smaller. As a conservative measure
of scattered light all light in a 5$^\circ$ diameter field of view of 
a Cherenkov telescope centered on a gamma source is counted here --
although in practice pixels more than 0.5$^\circ$ from the image
major axis would generally be below a minimum threshold and not
counted. Note that almost all light scattered by less than 2.5$^\circ$
has a delay of less than 10~ns with respect to direct light and short
integration times would not significantly suppress scattered light.

\begin{figure}
\hc{\psfig{file=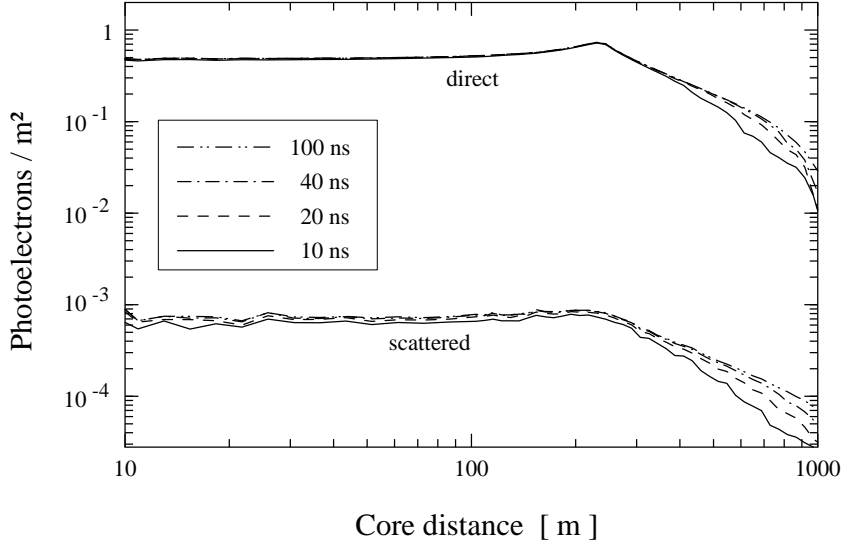,width=0.8\textwidth}}
\caption[Direct and scattered light at 60$^\circ$ zenith angle]{Lateral
distribution of direct and scattered light in 300~GeV $\gamma$-showers
at a zenith angle of 60$^\circ$ (average of 250 simulated showers). 
Core distances are given in a
plane perpendicular to the shower axis. A 5$^\circ$ diameter field of view
has been assumed, and mirror reflectivity and PM quantum efficiency as before.}
\label{fig:CT-scattered-60deg}
\end{figure}

In the small field of view of such telescopes the scattered
light has approximately the same path length as direct light and the
ratio of scattered to direct light approximately scales with the
airmass ($1/\cos z$ for a plane-parallel atmosphere). 
Since experimental groups
are more concerned about scattered light in large-zenith-angle observations
than near vertical, Figure~\ref{fig:CT-scattered-60deg} shows the case
for zenith angle $z=60^\circ$. Even in this case, scattered light is
quite marginal (of the order of $10^{-3}$). 

It should be noted, however, that some aerosol scattering
phase function models for the boundary layer (e.g.\ as in
MODTRAN, see Figure~\ref{fig:phase-functions}) 
have an additional peak in the very forward direction,
which is not present in the MODTRAN 2--10~km tropospheric scattering
model and not accounted for by the Henyey-Greenstein phase function.
Under such circumstances -- mainly Cherenkov experiments not far above
the surrounding terrain or observing in the presence of thin clouds or fog -- 
small-angle scattering of the Cherenkov
light could be up to an order of magnitude more severe. This, for example,
is the case in simulations of sea level observations with the MODTRAN
rural haze phase function for aerosol scattering in the lower 2~km, where
scattered light in the 5$^\circ$ field of view amounts to 1\% of direct light
at 60$^\circ$ zenith angle.
Even at that level, scattered light should not be of major concern.

The impact of thin clouds -- which has not been simulated here -- 
should be primarily on the trigger rate. Depending on the cloud
altitude, the image {\em length\/} parameter could be reduced by losing 
only light emitted above the layer but the impact of scattered light on image
parameters would still be small. Image parameters could rather be
affected by the change of night-sky noise -- which could either be reduced
due to absorption or increased due to scattering of urban light.


\begin{figure}
\hc{\psfig{file=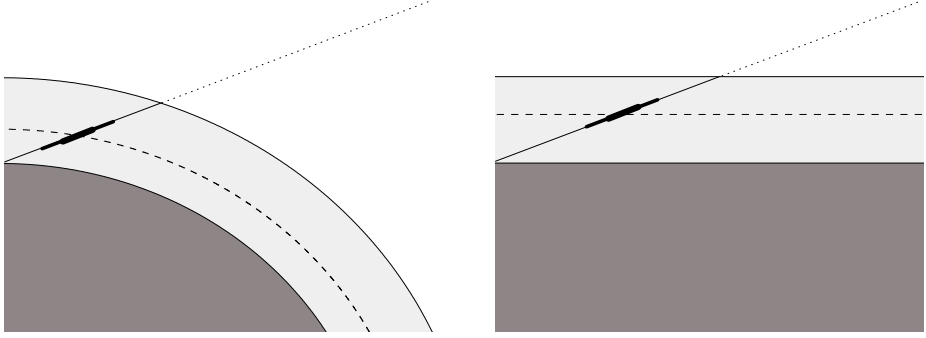,width=0.9\textwidth}}
\caption[Spherical versus planar atmosphere]{In the correct spherical
description of the atmosphere all slant shower paths have less material
to traverse than in the planar description. As a consequence the
shower maxima at fixed atmospheric depth are at lower altitudes 
(dashed lines) in the spherical description.}
\label{fig:sphere+plane}
\end{figure}

\begin{figure}
\hc{\psfig{file=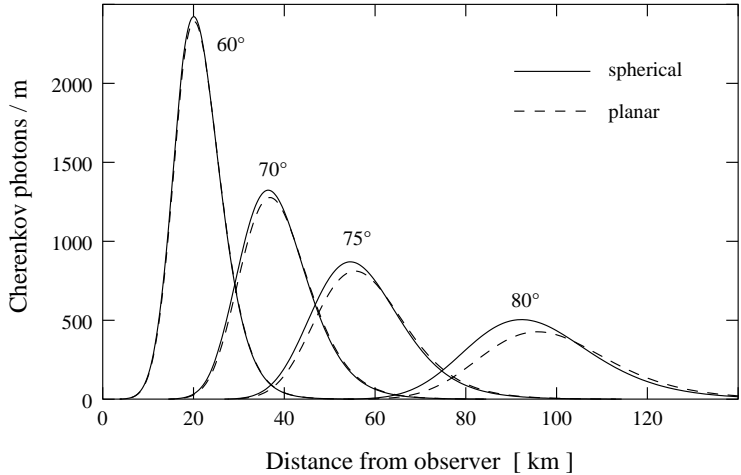,width=0.7\textwidth}}
\caption[Cherenkov emission for spherical and planar atmosphere]{The
average longitudinal Cherenkov emission profile as a function of
distance from the observer (Cherenkov photons emitted in the
wavelength range 300--600~nm per meter along the shower axis).
An analytical approximation is used for different
zenith angles (solid lines: spherical geometry,
dashed: planar geometry).}
\label{fig:sphere+plane_long}
\end{figure}

\section { Planar versus spherical atmosphere }

The CORSIKA program is at present using a plane-parallel atmosphere
for shower simulations -- except for a special, not Cherenkov-enabled,
{\em horizontal} version. The impact of a planar versus a spherical
atmosphere, in its qualitative implications, is nevertheless easy
to show (see Figure~\ref{fig:sphere+plane}). With spherical geometry
the shower maximum (at constant atmospheric depth) 
is at a lower altitude, where the index of refraction
is larger and more Cherenkov light is emitted.
This geometry effect is only relevant for large zenith angles.

Instead of shower simulations by the Monte Carlo method, an analytical
approximation of the average Cherenkov emission of gamma showers is
used here. It takes the longitudinal shower profile, the Cherenkov
emission threshold depending on the index of refraction, and the emission
of electrons above threshold into account. This approximation 
reproduces the longitudinal profile of Cherenkov emission in 
CORSIKA simulations very well for all model atmospheres.
This approximation has been used with both spherical and planar
atmospheric geometry to show that the difference is
insignificant below 60$^\circ$ zenith angle, and 
little significant below 70$^\circ$ (see Figure~\ref{fig:sphere+plane_long}).
For hadronic showers there is a small additional effect of fewer
pions and kaons decaying before the next interaction and, thus, fewer muons
with the spherical geometry at very large zenith angles.


\section { Refraction }

One of the recent achievements in VHE energy $\gamma$-astronomy
is the fact that TeV $\gamma$-ray sources can be located with
sub-arcminute accuracy \cite{puehlhofer}. In addition, observations
at large zenith angles are carried out by more and more Cherenkov
telescope experiments, either to extend the observation time
for a source or the effective area for high-energy showers,
or to detect sources only visible at large zenith angles.
Refraction of Cherenkov light in the atmosphere is therefore
of increasing concern but is usually either neglected entirely
or only considered in a qualitative way. The following discussion
is based on numerical ray-tracing. The refraction method built
into recent CORSIKA versions is based on a fit to such ray-tracing.

For a plane-parallel atmosphere
Snell's law of refraction is
\begin{equation}
n(z)\,\sin\theta(z) = \rm{const.}
\end{equation}
with $n(z)$ being the index of refraction at altitude $z$ and
$\theta(z)$ being the zenith angle of the ray at this altitude.
For a spherical atmosphere 
\begin{equation}
n(z)\,(R_{\rm{E}}+z)\,\sin\theta(z) = \rm{const.}
\end{equation}
has to be used instead, with $R_{\rm{E}}$ being the earth radius.

\begin{figure}
\hc{\psfig{file=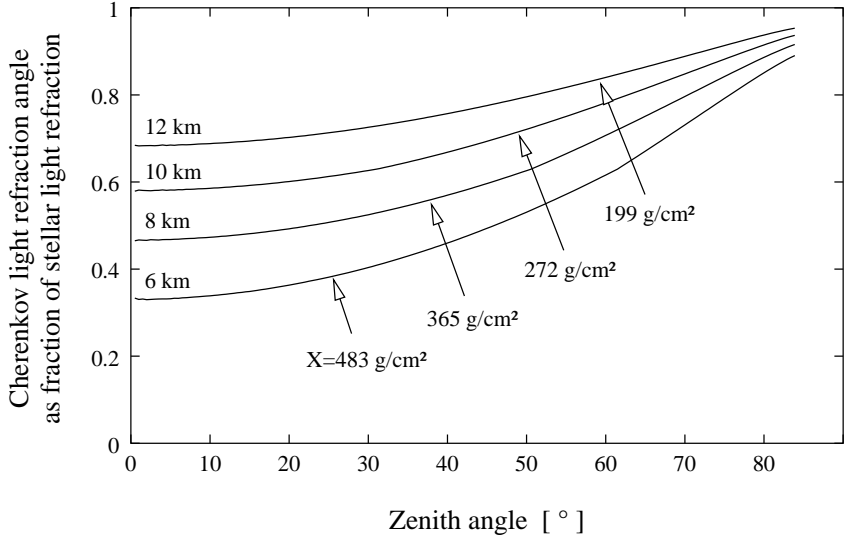,width=0.8\textwidth}}
\caption[Refraction angle of Cherenkov light]{The refraction angle of
Cherenkov light as a function of zenith angle, expressed as a fraction of
the corresponding refraction of stellar light. Numerical integrations
for U.S. standard atmosphere with spherical geometry and 
2200~m observation altitude.
Curves shown are for emission at constant atmospheric depth along 
the shower axis. For vertical showers, the depths of
483, 365, 272, and 199 g~cm$^{-2}$ correspond to
altitudes of 6, 8, 10, and 12~km, respectively.}
\label{fig:refract}
\end{figure}

The refraction of Cherenkov light emitted in the atmosphere is
evidently smaller than that of star light seen from the same
direction. Thus, even when using guide stars for tracking of
Cherenkov telescopes, a correction for refraction has to be
applied to take full advantage of measured shower directions.
For $\gamma$-showers of 0.1--1~TeV the Cherenkov light is
refracted typically 60--50\% (70--60\%) as much as 
stellar light up to 40$^\circ$
(near 60$^\circ$) zenith angle, with less refraction for showers
of higher energy (see Figure~\ref{fig:refract}). 
The different amount of refraction
of light from the beginning and the end of the shower, respectively,
leads to a change of image length. When an inclined shower is
seen from below the axis, it appears slightly shorter, and when seen
from above the axis, it appears longer -- by a fraction of an arcminute.

Apart from the change in the Cherenkov light direction, refraction
also affects the arrival position and time. The impact on the
arrival time is marginal if measured in a plane perpendicular
to the shower axis and is well below one nanosecond even at 80$^\circ$
zenith angle. The arrival position in the shower plane is affected
by typically 3~meters (18~m) in TeV $\gamma$-showers of 60$^\circ$ (75$^\circ$)
zenith angle. Actual changes depend on the height of emission and lead
to a small distortion of the lateral shape -- the average shift being
irrelevant. 


\section { Conclusions }

The impact of a number of atmospheric parameters on the atmospheric
Cherenkov technique in general have been presented. It turns out
that there are several such parameters which deserve more attention
in the experimental analysis. This applies to the imaging technique
in VHE $\gamma$-ray astronomy and to non-imaging techniques --
both in $\gamma$-ray and cosmic-ray studies. 

The vertical structure
of the atmosphere is the most striking of these parameters, with
up to 60\% difference in Cherenkov light between tropical and
polar models and some 15--20\% seasonal variation at mid-latitude
sites. The appropriate structure could be easily applied in
shower simulations for particular sites -- while most calculations
so far were restricted to US Standard Atmosphere 1976 or similar profiles.
Relevant measurements are routine procedure of many meteorological
institutions and data are readily available.

Atmospheric extinction of Cherenkov light appears as another area
which deserves perhaps more care, as the accuracy of the experiments
improves. In particular, the assumption that extinction (scattering
and absorption) by aerosols is, like Rayleigh scattering, only
a function of traversed atmospheric thickness, is not a very good one.
Extrapolation of high-altitude extinction measurements to
low-level sites must be avoided. Cherenkov experiments should,
at least, apply standard astronomical extinction 
monitoring procedures -- if not already available from co-located
optical observatories. In order to reduce energy systematics below
ten percent this might, however, not be enough.

Measurement of the aerosol
vertical structure is -- 
in contrast to stellar light extinction or to the air density profile --
rather difficult. Lidar remote sensing methods,
for example, measure primarily the back-scattered amount of light,
although lidar methods are available to measure also the extinction profile
with a reasonable 10\% accuracy.
The ratio of back-scattered light to total extinction
depends very much on aerosol composition. Even otherwise similar
phase functions (see Figure~\ref{fig:phase-functions}) differ easily
by a factor of two in the back-scattering regime.
As a consequence, some model-dependence will likely remain in
transmission calculations but the best available models
for a site should be used.

For UV Cherenkov experiments sensitive below 300~nm wavelength,
the variations of ozone profiles are an additional area of concern
and absorption on oxygen can no longer be ignored.

Scattered Cherenkov light has a rather minor impact for Cherenkov
experiments -- in contrast to air shower fluorescence experiments.
Scattering should, however, be of some concern to wide-angle
non-imaging experiments using the Cherenkov light lateral distribution
either to discriminate between hadronic and gamma showers or to
disentangle the cosmic-ray mass composition.

The assumption of a plane-parallel atmosphere in air shower simulations
becomes a problem when going to zenith angles beyond about $75^\circ$.
These large zenith angles are very important for fluorescence
experiments in order to achieve the largest possible effective areas.
For Cherenkov $\gamma$-ray experiments the increase in effective area,
however, will -- under most experimental conditions that can be envisaged --
be more than counterbalanced by the much worse gamma-hadron discrimination
when observing showers from more than 60~km distance. As a consequence,
implementation of the proper spherical geometry in shower simulations 
is of less importance to the Cherenkov than to the fluorescence method.

Atmospheric Cherenkov light refraction -- until recently negligible
compared to experimental errors -- has to be accounted for when locating
sources with sub-arcminute accuracy as now possible. For the time
being, an approximate correction, e.g. as taken from
Figure~\ref{fig:refract}, should generally be sufficient. 
For more accurate results, enhancements to the
CORSIKA code should be available with new CORSIKA versions.

In addition to those atmospheric parameters covered in this
paper, there are other, more subtle effects at work.
One such example would be the time-variable night-sky background,
e.g.\ due to airglow and scattering of urban illumination -- something
that should be kept in mind but is probably beyond the scope of what
should be or can be accurately modelled in shower simulations.

The geomagnetic field -- not covered by this paper --
has an additional impact. For present IACT experiments, say at
energies of the order of 1~TeV, the main impact is a variation
of the Cherenkov light intensity by a few percent between observations
parallel to and perpendicular to the field direction. Image
parameters are little affected at these energies. For future experiments
observing lower energy showers with better angular resolution
the geomagnetic field can be expected to be more significant.


\section*{Acknowledgements}
Radiosonde measurements for Tenerife were kindly provided by the
Izana Global Atmospheric Watch (GAW) Observatory. The use of
further radiosonde data from the Amundsen-Scott and Neumayer
stations is acknowledged, as well as the use of La Palma extinction 
data provided by the Isaac Newton group on La Palma, Canary Islands, and 
the Royal Greenwich Observatory, Cambridge. Most of this paper
is based on calculations with CORSIKA 5.7 and it is a pleasure to
thank its authors, in particular D.~Heck, for their support.
The MODTRAN 3~v1.5 program was kindly provided by the Phillips
Laboratory, Geophysics Directorate, at Hanscom AFB,
Massachusetts (USA).


\end{document}